\documentstyle[12pt,fleqn]{article}

\begin{document}

\title{Conformal Decomposition of the Effective Action
       and Covariant Curvature Expansion}

\author{ A. O. Barvinsky, A. G. Mirzabekian  \\
         \em Lebedev Physics Institute and
             Research Center in Physics  \\
         \em Leninsky Prospect 53, Moscow 117924, Russia  \\
         and  \\
         V. V. Zhytnikov  \\
         \em Moscow State Pedagogical University,  \\
         \em Davydovskii 4, Moscow 107140 and  \\
         \em Lebedev Research Center in Physics }

\date{}
\maketitle

\begin{abstract}
The class of effective actions exactly reproducing the
conformal anomaly in 4D is considered. It is demonstrated
that the freedom within this class can be fixed by the
choice of the conformal gauge.
The conformal invariant part of the generic one-loop effective
action expanded in the covariant series up to third order
in the curvature is rewritten in the new conformal basis.
The possible applications of the obtained results are
discussed.
\end{abstract}


\section{Introduction: 4D vs 2D}

This is a well-known fact that the local conformal invariance
is a powerful tool on the way to exactly solvable models in QFT.
The culmination point of this idea ia a 2-dimensional theory of
 massless conformal scalar field underlyning a huge building
of the modern string and conformal field theory. This theory
is famous due to the fact that its exact quantum effective
action is known in a closed form of the Polyakov action
\cite{Pol,FrolVilk}
\begin{equation}
W_{\rm P} = \frac{1}{2}{\rm Tr}\ln\Box =
\frac{1}{96\pi}\int d^2 x g^{1/2} R \frac{1}{\Box} R,
\end{equation}
that can be obtained by integrating the local
conformal anomaly
\begin{equation}
g_{\mu\nu}\frac{\delta W_{\rm P}}{\delta g_{\mu\nu}} =
-\frac{1}{48\pi} g^{1/2} R.
\end{equation}
The extension of this procedure to higher dimensions,
namely to 4D, is also known and exists in the form of
the Riegert action \cite{Riegert}
\begin{eqnarray}
W_{\rm R} &=& \frac{1}{2(4\pi)^2} \int d^4 x g^{1/2}
\left\{\frac{1}{4}\left[aC^2+dF^2
+\frac{b}{2}\left(E-\frac{2}{3}\Box R\right)\right]\right.
\nonumber\\[2mm]
&&\qquad \left.\times \frac{1}{\cal D}\left(E
-\frac{2}{3}\Box R\right)
-\left(\frac{c}{12}+\frac{b}{18}\right) R^2 \right\},
\label{Riegert}
\end{eqnarray}
which also generates the local conformal
anomaly in 4D:
\begin{equation}
g_{\mu\nu}\frac{\delta W_{\rm R}}{\delta g_{\mu\nu}} =
\frac{1}{4(4\pi)^2} g^{1/2}
\left(aC^2+bE+dF^2+c\Box R\right) \equiv \frac12 T_{\rm A},
\label{anoma}
\end{equation}
where
\[
C^2 = C^2_{\mu\nu\alpha\beta}, \quad
F^2 = F^2_{\mu\nu}, \quad
E = R^2_{\mu\nu\alpha\beta}-4R^2_{\mu\nu}+R^2,
\]
are respectively the squares of the Weyl tensor, the square of the
gauge field
strength and Euler density, $a,b,c$ and $d$ are numerical
constants specifying a concrete conformal-invariant model in 4D,
and $1/{\cal D}$ is an inverse (Green's function) of a special
4-th order differential operator
\begin{equation}
{\cal D} = \Box^2 + 2R^{\mu\nu} \nabla_\mu\nabla_\nu
-\frac{2}{3}R\Box+\frac{1}{3}(\nabla^\mu R)\nabla_\mu,
\label{Operator}
\end{equation}
having a property of beeing conformal invariant
when acting on a scalar field of zero conformal weight.
Here and below we use the sign conventions
$R^\mu_{\,\cdot\, \alpha\nu\beta}=\partial_\nu
\Gamma^\mu_{\alpha\beta}-\cdots,$  $R_{\alpha\beta}=
R^\mu_{\,\cdot\, \alpha\mu\beta},$
$R=g^{\alpha\beta}R_{\alpha\beta}$.

The both Polyakov and Riegert actions are remarkably
similar in structure. Apart from the local  $R^2$ term
they are both nonlocal, the Green's function of the
4-th order operator ${\cal D}$ beeing a 4D-generalization
of 2D covariant D'Alambertian $\Box$. Apart from the $C^2$
and $F^2$ terms, nonlocal Riegert action is quadratic in
the quantity $E-\frac23\Box R$, which is a density of the
topological invariant - total derivative term - the 4D
generalization  of the 2D Euler density $R$ :
\begin{equation}
R\frac{1}{\Box}R
\quad\longleftrightarrow\quad
\left(E-\frac{2}{3}\Box R\right)
\frac{1}{{\cal D}}
\left(E-\frac{2}{3}\Box R\right).
\end{equation}
This emphasizes the topological information encoded in
2D and 4D anomalous effective actions. Finally, even their
nonlocalities have a similar nature because a naive infrared
behaviour of their Green functions is logarithmic
\[
\frac{1}{\Box}\delta(x,y) \sim
\frac{1}{\cal D}\delta(x,y) \sim
\ln\frac{|x-y|}{\mu},
\]
and strictly speaking, does not make sense in
asymptotically-flat spacetime, unless the Green's functions
are acting upon the total derivative terms \cite{I} (which is
just the case of both actions).

The above list of similarities may not be complete, but there
exists an essential difference between two and four dimensions:
while the one-loop action in 2D is actually exhausted by the
Polyakov action (up to inessential constant), the Riegert action
in 4D represents only anomalous part of the total action
defined up to a nontrivial conformal invariant functional
$\overline{W}$ of the metric and matter fields, bearing
important physical information:
\begin{equation}
W_{\rm 1-loop}^{\rm(2D)} = W_{\rm P}+{\rm const}
\quad\longleftrightarrow\quad
W_{\rm 1-loop}^{\rm(4D)} = W_{\rm R}+\overline{W}.
\end{equation}
Such a conformal decomposition of the full action is,
obviously not unique and very little is known about its
conformal invariant part $\overline{W}$. So, the purpose of
our work is:
\begin{itemize}
\item[(i)] to find a class of conformal decompositions of the
one-loop effective action and show that the freedom within
this class can be fixed by the choice of the conformal gauge;
\item[(ii)] as a by-product of the above, to find a class
of non-local effective actions generating the conformal
anomaly and establish the status of Riegert's action in
this class;
\item[(iii)] to calculate the conformal invariant part
of the effective action $\overline{W}$ by the technique of the
covariant curvature expansion and demonstrate the drastic
simplification of the latter in the new conformal basis
of the non-local invariants;
\item[(iv)] briefly discuss and speculate on the applications
of this conformal decomposition in black-hole physics,
theory of the quantum gravitational collapse and possible
extension of the 2D methods of the conformal field theory
to higher dimensions.
\end{itemize}

\section{Integrating the conformal anomaly}

Given the conformal anomaly $T_{\rm A}(g)$, the equation
\begin{equation}
g_{\mu\nu}\frac{\delta W_{\rm R}}{\delta g_{\mu\nu}} =
\frac12 T_{\rm A},
\end{equation}
can be easily integrated along the orbit of the local
conformal group
\begin{equation}
g_{\mu\nu}(x) = e^{-\sigma(x)}\overline{g}_{\mu\nu}(x),
\label{cg}
\end{equation}
to give
\begin{equation}
W[g] = W[\overline{g}] + \Delta W[\sigma,\overline{g}], \qquad
\Delta W[\sigma,\overline{g}] = -\frac12\int^\sigma_0 d \sigma'
T_{\rm A}(e^{-\sigma'}\overline{g}).\label{x}
\end{equation}
Omitting the technical details, let us mention that the
resulting $\Delta W[\sigma,\overline{g}]$ is quartic in
the conformal group parameter $\sigma(x)$ and its derivatives
\[
\Delta W \sim \sigma+\sigma^2+\sigma^3+\sigma^4.
\]
It is remarkable, however, that cubic and quartic terms here
can be absorbed into a simple conformal invariant
$\int d^4x \overline{g}^{1/2} R^2(\overline{g})$ -- the
functional of $\overline{g}_{\mu\nu}$, so that $\Delta W$
takes the form
\begin{eqnarray}
\Delta W &=& \frac{1}{2(4\pi)^2} \int d^4 x g^{1/2}
\left\{-\frac12\left[aC^2+dF^2
+b\left(E-\frac{2}{3}\Box R\right)\right]\sigma\right.
\nonumber\\[2mm]
&&\qquad\qquad\qquad \left.
-\frac{b}{2}\sigma{\cal D}\sigma-\left(\frac{c}{12}+\frac{b}{18}
\right)R^2\right\} \nonumber\\[2mm]
&&\qquad +\frac{1}{2(4\pi)^2} \frac{b}{18}
\int d^4x \overline{g}^{1/2} R^2(\overline{g}),
\label{xx}
\end{eqnarray}
where the quadratic in $\sigma$ form contains the 4-th
order conformal invariant operator ${\cal D}$ (\ref{Operator}).

\section{Conformal gauge fixing}

The further manipulations with eqs. (\ref{x}) and (\ref{xx})
are aimed to declare $\overline{g}_{\mu\nu}(x)$ to be the
conformal invariant -- the unique representative of the
class of equivalence of metrics belonging to the orbit of the
conformal group (\ref{cg}). This can be attained by a conventional
gauge-fixing procedure: choosing the surface of some
conformal gauge conditions
\begin{equation}
\chi(g) =0,
\end{equation}
transversal to the conformal orbit and having with this orbit
the only one intersection point $\overline{g}$.
This configuration-space
point $\overline{g}=ge^{\sigma}$, when substituted into the
gauge condition, results in the equation for conformal gauge
parameter
\begin{equation}
\chi(ge^{\sigma})=0\quad \Rightarrow\quad
\sigma=\Sigma(g),
\end{equation}
shifting  arbitrary metric $g_{\mu\nu}(x)$ along its
conformal orbit to its conformal invariant representative
\begin{equation}
\overline{g}_{\mu\nu} = e^{\Sigma(g)}g_{\mu\nu},
\qquad g_{\alpha\beta} \frac{\delta}{\delta g_{\alpha\beta}}
\overline{g}_{\mu\nu}[g]=0.
\label{barmetric}
\end{equation}
Thus the needed conformal decomposition takes the form
\begin{eqnarray}
W[g] &=& \overline{W}[g] + W_{\rm A} [g,\Sigma(g)], \nonumber\\[2mm]
\overline{W}[g] &=& W[\overline{g}]
+\frac{1}{2(4\pi)^2} \frac{b}{18}
 \int d^4x \overline{g}^{1/2} R^2(\overline{g}),\quad
\overline{g}=e^{\Sigma(g)}g, \nonumber\\[2mm]
W_{\rm A}[g,\Sigma] &=& \frac{1}{2(4\pi)^2} \int d^4 x g^{1/2}
\left\{-\frac12\left[aC^2+dF^2
+b\left(E-\frac{2}{3}\Box R\right)\right]\Sigma\right.
\nonumber\\[2mm]
&&\quad \left.
-\frac{b}{2}\Sigma{\cal D}\Sigma-\left(\frac{c}{12}+\frac{b}{18}
\right)R^2\right\}
\label{xxx}\label{WA}
\end{eqnarray}
where $\overline{W}[g]$ and $W_{\rm A}[g,\Sigma(g)]$
are its conformal-invariant and anomalous parts respectively.

There are two distinguished, so to say ``exactly solvable''
conformal gauges in 4D, in which the gauge parameter
$\Sigma(g)$ can be calculated in a closed form as a functional
of $g$. One is due to Fradkin and Vilkovisky \cite{FradVilk}
\begin{equation}
\chi_{\rm FV}(g) = R(g),
\end{equation}
with
\begin{equation}
\Sigma_{\rm FV}(g) = 2\ln\left(1+\frac{1}{6}
\frac{1}{\Box-R/6} R\right),
\label{SigmaFV}
\end{equation}
the operator $\Box-R/6$ (conformal covariant when acting on a
scalar of $-1$ conformal weight) playing the role of the
corresponding Faddeev--Popov operator. Another gauge is
intrinsically associated with the structure of $\Delta W$
in (\ref{xx}), and can be called the Riegert gauge \cite{Riegert}
\begin{equation}
\chi_{\rm R} (g) = E(g) - \frac23 \Box R(g),
\end{equation}
for which
\begin{equation}
\Sigma_{\rm R}(g) = -\frac1{2{\cal D}}
\left(E-\frac23 \Box R\right),
\label{SigmaR}
\end{equation}
and the anomalous part of (\ref{xxx}) coincides with
the Riegert one (\ref{Riegert}) :
\[
W_{\rm A} [g,\Sigma_{\rm R}(g)] = W_{\rm R}[g].
\]

Curiously enough, there is only one analogous ``exactly solvable''
conformal gauge in 2D, $R(g)=0$ with $\Sigma(g)=(1/\Box) R$.
It would be interesting to check if the hierarchy of such
distinguished gauges grows with the spacetime dimensionality.

\section{Conformal invariant curvature expansion for
$\overline{W}$}

Very little is known about the conformal invariant
part of the effective action $\overline{W}$, and, to the
best of our knowledge, it can be calculated for a generic
background only within certain approximation schemes.
One such scheme recently developed in a series of papers
\cite{I,II,III,IV} represents a covariant expansion in powers
of curvatures and field strengths with the coefficients -
the nonlocal operator form factors. The interest in this
particular approximation  arises from a simple model
of the quantum gravitational collapse of a spherical thin
null shell considered by Frolov and Vilkovisky \cite{FrolVilk}.
Under very weak assumptions they showed that for quantum
effective action modelling the highest order derivatives
\[
W = \int d^4x\,g^{1/2}(R-a C^2_{\mu\nu\alpha\beta}) + \dots
\]
(with some nonperturbative constant $a$) the collapsing
null shell of a small mass
${\cal M} \ll \sqrt{a}$
does not develop either horizon or singularity,
and for large ${\cal M}$ is likely to generate a spacetime
geometry which is free from singularities and has a closed
apparent horizon surrounding the domain of large but finite
spacetime curvature uniformly bounding the curvature
throughout the whole spacetime. This brings to life the idea
of expansion in powers of curvatures and field strengths of
small magnitude but however rapidly varying in spacetime.

For the one-loop effective action
\begin{equation}
W=\frac 12\,{\rm Tr \,ln}\,H,    \label{WH}
\end{equation}
in Euclidean noncompact asymptotically flat spacetime the expansion
up to third order in the curvature $\Re$ looks as follows \cite{IV}:
\begin{eqnarray}
-W&=&{\frac1{2(4\pi)^2}\int\! d^4x\, g^{1/2}\, \,{\rm tr}\,}
\left\{\sum^{5}_{i=1}\gamma_i(-\Box_2)\Re_1\Re_2({i})\right.
\nonumber\\
&&\qquad\qquad\left.+\sum^{29}_{i=1}
\Gamma_{i}(-\Box_1,-\Box_2,-\Box_3)
\Re_1\Re_2\Re_3({i})\right\}+{\rm O}[\Re^4].
\label{result3}
\end{eqnarray}
Here $H$ is a generic second order minimal operator
\begin{equation}
H=\Box \hat1 + \Big(\hat P-\frac16\,R\hat 1\,\Big),\qquad
\Box=g^{\mu\nu}\nabla_{\mu}\nabla_{\nu}, \label{oper}
\end{equation}
acting on an arbitrary set of fields $\varphi^A(x)$.
Here $A$ stands for any set of discrete indices,
and the hat indicates that the quantity is a matrix in the
corresponding vector space of field components.
The operator (\ref{oper}) depends on three arbitrary
quantities: spacetime metric $g_{\mu\nu}$, covariant
derivative connection and a potential $\hat P$.

The quadratic $\Re_1\Re_2(i)$ and cubic $\Re_1\Re_2\Re_3(i)$
structures in (\ref{result3}) are invariants which are constructed
with the help of the corresponding
three ``curvatures'' \cite{II,IV,Basis}:
\begin{equation}
\Re=\{ R_{\mu\nu},\;\hat{\cal R}_{\mu\nu},\;\hat P \},
\label{curvs}
\end{equation}
where $\hat{\cal R}_{\mu\nu}$ is a commutator curvature
$[\nabla_\mu,\nabla_\nu]\varphi
= \hat{\cal R}_{\mu\nu}\varphi$.
An important feature of the covariant curvature expansion is
that it uses Ricci curvature tensor rather than the Riemann one
due to the fact that in the asymptotically flat spacetime
the Riemann tensor can always be eliminated via the
differentiated Bianchi identity
\begin{eqnarray}
R^{\alpha\beta\mu\nu} &=& \frac1{\Box}\left(
 \nabla^\mu \nabla^\alpha R^{\nu\beta}
+\nabla^\nu \nabla^\beta R^{\mu\alpha}
-\nabla^\nu \nabla^\alpha R^{\mu\beta}
-\nabla^\mu \nabla^\beta  R^{\nu\alpha}
\right)
\nonumber\\&&\qquad
+{\rm O}[R^2].
\label{RimRic}
\end{eqnarray}

The form factors $\gamma_i,\Gamma_i$ in the covariant
curvature expansion (\ref{result3}) are nonlocal operator
functions of d'Alambertian operator $\Box$ (subscripts
$1,2,3$ mean that $\Box_1$ acts on $\Re_1$ only
and similarly for $\Box_2,\Box_3$).
The explicit expressions for the second order form factors
$\gamma_i$ were obtained in \cite{II} and have a rather
simple structure
\begin{equation}
\gamma_i(-\Box) = {\rm const}\cdot\ln(-\Box/\mu^2)+{\rm const}.
\end{equation}

The third order form factors $\Gamma_i$ are
significantly more complicated and schematically
can be represented as follows
\begin{eqnarray}
&& \Gamma_i(-\Box_1,-\Box_2,-\Box_3) = A\,
   \Gamma(-\Box_1,-\Box_2,-\Box_3) \nonumber\\
&&\qquad\qquad\qquad\qquad\qquad\qquad +
\sum_{i,k=1; i<k}^3C_{ik}\frac{\ln(\Box_i/\Box_k)}
{(\Box_i-\Box_k)} + B,
\label{ff3}
\end{eqnarray}
where $A,B,C_{ik}$ are complicated rational functions of
$\Box_1,\Box_2,\Box_3$. The {\em fundamental form factor}
$\Gamma(-\Box_1,-\Box_2,-\Box_3)$ can not be expressed
in elementary functions but has several integral
representations \cite{IV}. The actual expressions for
(\ref{ff3}) are huge and take pages \cite{IV}.

If one considers the operator $H$ which corresponds to
some conformal invariant theory, say complex conformal
scalar field with $\lambda\phi^4$ interaction, it is
possible to obtain the anomaly (\ref{anoma})
by direct variation of (\ref{result3}) \cite{Anom}.
Unfortunately this derivation has an important drawback:
the resulting anomaly is not exact for the quantity
$R^2_{\mu\nu\alpha\beta}$ is expanded in powers of
Ricci tensor due to (\ref{RimRic}). The effective
action (\ref{result3}) reproduces anomaly only up to
second order in the curvature inclusive.

Another important observation comes from the fact that the
anomaly (\ref{anoma}) is local while the conformal variation
of each $\Re_1\Re_2\Re_3(i)$ produces a contribution containing
the essentially nonlocal fundamental
form factor $\Gamma$. In the final answer all terms
with $\Gamma$ cancel each other, which was verified by
the direct computation \cite{Anom}. This means that not
all form factors $\Gamma_i$ are actually independent
but some of them are linear combinations of the others.

Our goal is to utilize these facts to rewrite (\ref{result3})
in a more compact form accounting for the conformal anomaly.
In order to do so we are going to perform the conformal
decomposition of the action (\ref{result3}) on the anomalous
$W_{\rm A}$ and the conformal invariant part $\overline{W}$.
For sake of clarity, we restrict ourselves initially
to the purely gravitational part of the action which includes
ten cubic structures composed of spacetime curvatures
$R_{\mu\nu}$ and $R$ only:
\begin{eqnarray}
\Re_1\Re_2\Re_3({9})&=&R_1 R_2 R_3, \nonumber\\[1.5mm]
\Re_1\Re_2\Re_3({10})&=&R_{1\,\alpha}^\mu
   R_{2\,\beta}^{\alpha} R_{3\,\mu}^\beta, \nonumber\\[1.5mm]
\Re_1\Re_2\Re_3({11})&=&
   R_1^{\mu\nu}R_{2\,\mu\nu}R_3, \nonumber\\[1.5mm]
\Re_1\Re_2\Re_3({22})&=&R_1^{\alpha\beta}
   \nabla_\alpha R_2 \nabla_\beta R_3,\nonumber\\[1.5mm]
\Re_1\Re_2\Re_3({23})&=&\nabla^\mu R_1^{\nu\alpha}
   \nabla_\nu R_{2\,\mu\alpha}R_3,\nonumber\\[1.5mm]
\Re_1\Re_2\Re_3({24})&=&R_1^{\mu\nu}
   \nabla_\mu R_2^{\alpha\beta}
   \nabla_\nu R_{3\,\alpha\beta},\nonumber\\[1.5mm]
\Re_1\Re_2\Re_3({25})&=&R_1^{\mu\nu}
   \nabla_\alpha R_{2\,\beta\mu}\nabla^\beta
   R_{3\,\nu}^\alpha, \nonumber\\[1.5mm]
\Re_1\Re_2\Re_3({27})&=&\nabla_\alpha\nabla_\beta
   R_1^{\mu\nu}\nabla_\mu\nabla_\nu R_2^{\alpha\beta}
   R_3,\nonumber\\[1.5mm]
\Re_1\Re_2\Re_3({28})&=&\nabla_\mu
   R_1^{\alpha\lambda} \nabla_\nu
   R_{2\,\lambda}^\beta\nabla_\alpha\nabla_\beta
   R_3^{\mu\nu}, \nonumber\\[1.5mm]
\Re_1\Re_2\Re_3({29})&=&\nabla_\lambda\nabla_\sigma
   R_1^{\alpha\beta}\nabla_\alpha\nabla_\beta
   R_2^{\mu\nu}\nabla_\mu\nabla_
   \nu R_3^{\lambda\sigma}.
\label{gbasis}
\end{eqnarray}

We can rewrite this part of the effective action in a new
basis $\overline{\Re_1\Re_2\Re_3}(i)$ which is identical
to (\ref{gbasis}) but with Ricci tensor replaced by the
tensor $C_{\mu\nu}$:
\begin{equation}
R_{\mu\nu} \longrightarrow C_{\mu\nu},
\end{equation}
which is a nonlocal contraction of the Weyl tensor
\begin{equation}
C_{\mu\nu}
\ \stackrel{\rm def}{=}\
\frac{2}{\Box}\nabla^\pi\nabla^\tau
C_{\mu\pi\nu\tau}.
\end{equation}
The relationship between $R_{\mu\nu}$ and $C_{\mu\nu}$ is
\begin{equation}
C_{\mu\nu} =
R_{\mu\nu}-\frac{1}{6}g_{\mu\nu}R
-\frac{1}{3}\frac{1}{\Box}\nabla_\mu\nabla_\nu R
+{\rm O}[R^2].
\label{RtoC}
\end{equation}
It is not hard to see that $C_{\mu\nu}$ is actually
a transverse-traceless part of the Ricci tensor
\[
C^\pi_\pi=0,\qquad \nabla_\pi C^\pi_\alpha={\rm O}[R^2].
\]

But the most important feature of $C_{\mu\nu}$ is that
in some sense it is a conformal invariant part of $R_{\mu\nu}$.
Indeed, for both conformal gauges (\ref{SigmaFV}),
(\ref{SigmaR})
\begin{equation}
R_{\mu\nu}(\overline{g}) =
R_{\mu\nu}(ge^{\Sigma(g)}) =
C_{\mu\nu}(g) + {\rm O}[R^2].
\end{equation}
The quantity $C_{\mu\nu}$ is not exactly conformal
invariant but unlike $R_{\mu\nu}$ the conformal
transformation of $C_{\mu\nu}$ is quadratic in curvature
\begin{equation}
C_{\mu\nu}(ge^{\sigma}) = C_{\mu\nu}(g)+{\rm O}[R^2].
\label{yyy}
\end{equation}

Taking into account (\ref{RtoC}) it is not hard to rewrite
the cubic part of the effective action (\ref{result3}) in the new
($C_{\mu\nu}$, $R$)-basis and compute the corresponding new
form factors. But now cubic structures can be subdivided
into two qualitatively different groups. The first group
includes five structures which consist of $C_{\mu\nu}$ only:
\begin{eqnarray}
\overline{\Re_1\Re_2\Re_3}({10})&=&C_{1\,\alpha}^\mu
   C_{2\,\beta}^{\alpha} C_{3\,\mu}^\beta, \nonumber\\[1.5mm]
\overline{\Re_1\Re_2\Re_3}({24})&=&C_1^{\mu\nu}
   \nabla_\mu C_2^{\alpha\beta}
   \nabla_\nu C_{3\,\alpha\beta},\nonumber\\[1.5mm]
\overline{\Re_1\Re_2\Re_3}({25})&=&C_1^{\mu\nu}
   \nabla_\alpha C_{2\,\beta\mu}\nabla^\beta
   C_{3\,\nu}^\alpha, \nonumber\\[1.5mm]
\overline{\Re_1\Re_2\Re_3}({28})&=&\nabla_\mu
   C_1^{\alpha\lambda} \nabla_\nu
   C_{2\,\lambda}^\beta\nabla_\alpha\nabla_\beta
   C_3^{\mu\nu}, \nonumber\\[1.5mm]
\overline{\Re_1\Re_2\Re_3}({29})&=&\nabla_\lambda\nabla_\sigma
   C_1^{\alpha\beta}\nabla_\alpha\nabla_\beta
   C_2^{\mu\nu}\nabla_\mu\nabla_
   \nu C_3^{\lambda\sigma}.
\label{CCCbasis}
\end{eqnarray}
Due to (\ref{yyy}) these structures do not contribute
to the conformal anomaly in a given approximation.
It is worth noting that form factors of the structures
(\ref{CCCbasis}) remain {\em exactly} equal to the form factors
of the corresponding original structures
$\Re_1\Re_2\Re_3(i),$ $i=10,24,25,28,29$.

Other five structures contain at least one scalar curvature:
\begin{eqnarray}
\overline{\Re_1\Re_2\Re_3}({9})&=&R_1 R_2 R_3, \nonumber\\[1.5mm]
\overline{\Re_1\Re_2\Re_3}({11})&=&
   C_1^{\mu\nu}C_{2\,\mu\nu}R_3, \nonumber\\[1.5mm]
\overline{\Re_1\Re_2\Re_3}({22})&=&C_1^{\alpha\beta}
   \nabla_\alpha R_2 \nabla_\beta R_3,\nonumber\\[1.5mm]
\overline{\Re_1\Re_2\Re_3}({23})&=&\nabla^\mu C_1^{\nu\alpha}
   \nabla_\nu C_{2\,\mu\alpha}R_3,\nonumber\\[1.5mm]
\overline{\Re_1\Re_2\Re_3}({27})&=&\nabla_\alpha\nabla_\beta
   C_1^{\mu\nu}\nabla_\mu\nabla_\nu C_2^{\alpha\beta} R_3.
\label{CCRbasis}
\end{eqnarray}
All these structures contribute to the anomaly
via conformal variation of $R$. But the form factors
of theese structures are subject to great simplification.
Instead of having the form (\ref{ff3}) they all boil
down to very short expressions whose structure can be
schematically represented as follows
\begin{equation}
\overline{\Gamma}_i =
\frac{1}{\Box} + \frac{\Box}{\Box}
\frac{\ln(\Box/\Box)}{(\Box-\Box)},
\end{equation}
where $1/\Box$ and $\Box/\Box$ stand for some simple
ratios of d'Alambertian operators like
$1/\Box_1$ or $\Box_1/\Box_2\Box_3$ etc.

Already on this stage we achieved our first goal --
the resulting expression for cubic part of the action is
significantly shorter than the original one.
But it is possible to push this simplification
even further. The idea is to eliminate the contribution of
structures (\ref{CCRbasis}) from the cubic part of the action
{\em completely}. It is possible to do this by including in
the action the exact anomalous part (\ref{WA}) and
replacing the rest of quadratic in curvature terms
by the explicitly conformal invariant functional.

A simple generalization of the anomalous action (\ref{WA})
to the case of the conformal invariant theory with the
inverse propagator (\ref{oper}) looks as follows. Let the operator
(\ref{oper}) be conformal covariant on the set of fields $\varphi^{A}(x)$
with some conformal weights, under the following transformations of
the metric and matter field strengths
\begin{equation}
g'_{\mu\nu} = e^{\sigma} g_{\mu\nu},\,\,
\hat{P}' = e^{\sigma} \hat{P},\,\,
\hat{\cal R}'_{\mu\nu}=
\hat{\cal R}_{\mu\nu}.   \label{conftr}
\end{equation}
Then the conformal anomaly for the action (\ref{WH}) is given by the
trace of the second Schwinger-DeWitt coefficient
\begin{equation}
\left( g_{\mu\nu}\frac{\delta}{\delta g_{\mu\nu}}+...\right) W =
-\frac{1}{2(4\pi)^2} g^{1/2} {\rm tr}\,\hat{a}_2(x,x),
\end{equation}
\begin{equation}
\hat{a}_2(x,x)=\frac16 \Box\hat{P}+\frac1{180}\Box R\hat{1}+
\frac 1{180}(R^2_{\mu\nu\alpha\beta}-R^2_{\mu\nu})\,\hat{1}+
\frac{1}{12}\hat{\cal R}^2_{\mu\nu}+\frac{1}{2}\hat{P}^2,
\end{equation}
where the dots denote the conformal variation with respect to relevant
matter fields participating in the transformations (\ref{conftr}).
One can easily check that this generic conformal anomaly can be
obtained from its particular case (\ref{anoma}) by replacing the
numerical parameters
$a,\,b$ and $c$ with $-{\rm tr}\,\hat{1}/60,\,{\rm tr}\,\hat{1}/180$ and
$-{\rm tr}\,\hat{1}/90$ respectively, substituting
$-\hat{\cal R}^2_{\mu\nu}/12-\hat{P}^2/2$ instead of $dF^2_{\mu\nu}/2$ and
adding the $\Box \hat{P}$ term. The corresponding anomalous action
can be obtained by the same modifications from (\ref{WA}) to give
\begin{eqnarray}
&&-W_{\rm A} = \frac{1}{2(4\pi)^2} \int d^4 x g^{1/2}
\,{\rm tr}\,\left\{
-\frac{1}{18}\hat{P}R
-\frac{\hat1}{1620} R^2
\right.\nonumber\\[2mm]&&\quad \left.
+\left[-\frac{\hat1}{120}C^2_{\mu\nu\alpha\beta}
  -\frac{1}{12}\hat{\cal R}^2_{\mu\nu}
  -\frac{1}{2}\hat{P}^2
  +\frac{\hat1}{360}\left(E-\frac{2}{3}\Box R\right)\right]
  \Sigma
\right.\nonumber\\[2mm]&&\quad \left.
+\frac{\hat1}{360}\Sigma{\cal D}\Sigma
\right\},
\label{Aresult}
\end{eqnarray}
where the local $\hat{P}R$ term generates the $\Box\hat{P}$ term of
the generic anomaly (note that this local finite term has been first
obtained in the second order of the covariant curvature expansion in
\cite{II}).

With this anomalous part the conformal decomposition of the effective
action (\ref{WH}) takes in Fradkin--Vilkovisky (\ref{SigmaFV}) or Riegert
(\ref{SigmaR}) gauges the form:
\begin{eqnarray}
 W &=& \overline{W} + W_{\rm A}.\label{Tresult}
\end{eqnarray}
Notice that, in our approximation for $\overline{W}$ cubic in the
curvatures, these two gauges are equivalent since
$\Sigma_{\rm FV}(g) \approx \Sigma_{\rm R}(g)=
\frac{1}{3}\frac{1}{\Box}R + {\rm O}[R^2]$
and $W_{\rm FV}$ and $W_{\rm R}$ differ only by
${\rm O}[R^4]$ terms.

The conformal invariant part of the action $\overline{W}$ reads
\begin{eqnarray}
-\overline{W}&=&
{\frac1{2(4\pi)^2}\int\! d^4x\, g^{1/2} \, \,{\rm tr}\,}
\left\{
-\frac{\hat1}{60} \overline{R}^{\alpha\beta}
\left(\ln(-\overline{\vphantom{I}\Box}_R/\mu^2)-\frac{16}{15}\right)
\overline{R}_{\alpha\beta}
\right.\nonumber\\[1.5mm]&&\quad
-\frac{1}{12} \hat{\cal R}^{\mu\nu}
\left(\ln(-\overline{\vphantom{I}\Box}_{\cal R}/\mu^2)-\frac{2}{3}\right)
\hat{\cal R}_{\mu\nu}
-\frac{1}{2} \hat{P}
\ln(-\overline{\vphantom{I}\Box}_P/\mu^2)
\hat{P}
\nonumber\\[1.5mm]
&&\quad \left.
+\sum
\overline{\Gamma}_{i}(-\Box_1,-\Box_2,-\Box_3)
\overline{\Re_1\Re_2\Re_3}({i})\right\}.
\label{Cresult}
\end{eqnarray}
Here
\begin{eqnarray}
\overline{R}_{\alpha\beta}(g) &=& R_{\alpha\beta}(\overline{g})
\nonumber\\
  & \equiv &
   R_{\alpha\beta} - \nabla_\alpha\nabla_\beta \Sigma
   -\frac{1}{2}g_{\alpha\beta}(\Box \Sigma + (\nabla\Sigma)^2)
   +\frac{1}{2}\nabla_\alpha\Sigma\nabla_\beta\Sigma \nonumber\\
 &=& C_{\alpha\beta} + {\rm O}[R^2],
\end{eqnarray}
and the operators
$\overline{\vphantom{I}\Box}_R$,
$\overline{\vphantom{I}\Box}_P$,
$\overline{\vphantom{I}\Box}_{\cal R}$
are defined in a such a way as to make the corresponding
$\overline{R}^2_{\alpha\beta}$, $\hat{P}^2$,
$\hat{\cal R}^2_{\mu\nu}$
terms {\em exactly} conformal invariant (and accounting for zero conformal
weights of covariant tensors $R_{\alpha\beta},\,\hat{\cal R}_{\mu\nu}$ and
the conformal weight $+2$ of $\hat {P}$):
\begin{equation}
\overline{\vphantom{I}\Box}_R(g)R_{\alpha\beta}
= \Box(\overline{g})R_{\alpha\beta}
= \Box(ge^{\Sigma(g)})R_{\alpha\beta},
\end{equation}
\begin{equation}
\overline{\vphantom{I}\Box}_{\cal R}(g)\hat {\cal R}_{\alpha\beta}
= \Box(\overline{g})\hat {\cal R}_{\alpha\beta},
\end{equation}
\begin{equation}
\overline{\vphantom{I}\Box}_P(g)\hat {P}
= e^{-\Sigma(g)}\Box(\overline{g})\left(e^{\Sigma(g)}\hat {P}\right).
\end{equation}

The sum in the cubic part of the action (\ref{Cresult})
includes only 19 structures, namely 5 gravitational
(\ref{CCCbasis}) and 14 matter-field ones:
\begin{eqnarray}
\overline{\Re_1\Re_2\Re_3}({1})
  &=& \hat{P}_1 \hat{P}_2 \hat{P}_3,\nonumber\\[1.5mm]
\overline{\Re_1\Re_2\Re_3}({2})
  &=& \hat{\cal R}^{\ \mu}_{1\ \alpha}
      \hat{\cal R}^{\ \alpha}_{2\ \beta}
      \hat{\cal R}^{\ \beta}_{3\ \mu},\nonumber\\[1.5mm]
\overline{\Re_1\Re_2\Re_3}({3})
  &=& \hat{\cal R}^{\mu\nu}_1
      \hat{\cal R}_{2\,\mu\nu}
      \hat{P}_3,\nonumber\\[1.5mm]
\overline{\Re_1\Re_2\Re_3}({5})
  &=& C_1^{\mu\nu}
      C_{2\,\mu\nu}
      \hat{P}_3,\nonumber\\[1.5mm]
\overline{\Re_1\Re_2\Re_3}({8})
  &=& C_1^{\alpha\beta}
      \hat{\cal R}_{2\,\alpha}^{\ \ \ \,\mu}
      \hat{\cal R}_{3\,\beta\mu},\nonumber\\[1.5mm]
\overline{\Re_1\Re_2\Re_3}({12})
  &=& \hat{\cal R}_1^{\alpha\beta}\nabla^\mu
      \hat{\cal R}_{2\mu\alpha}\nabla^\nu
      \hat{\cal R}_{3\nu\beta},\nonumber\\[1.5mm]
\overline{\Re_1\Re_2\Re_3}({13})
  &=& \hat{\cal R}_1^{\mu\nu} \nabla_\mu
      \hat{P}_2\nabla_\nu
      \hat{P}_3,\nonumber\\[1.5mm]
\overline{\Re_1\Re_2\Re_3}({14})
  &=& \nabla_\mu \hat{\cal R}_1^{\mu\alpha}\nabla^\nu
      \hat{\cal R}_{2\,\nu\alpha}
      \hat{P}_3,\nonumber\\[1.5mm]
\overline{\Re_1\Re_2\Re_3}({16})
  &=& \nabla^\mu C_1^{\nu\alpha}
      \nabla_\nu C_{2\,\mu\alpha}
      \hat{P}_3,\nonumber\\[1.5mm]
\overline{\Re_1\Re_2\Re_3}({17})
  &=& C_1^{\mu\nu}\nabla_\mu\nabla_\nu
      \hat{P}_2
      \hat{P}_3,\nonumber\\[1.5mm]
\overline{\Re_1\Re_2\Re_3}({18})
  &=& C_{1\,\alpha\beta}\nabla_\mu
      \hat{\cal R}_2^{\mu\alpha}\nabla_\nu
      \hat{\cal R}_3^{\nu\beta},\nonumber\\[1.5mm]
\overline{\Re_1\Re_2\Re_3}({19})
  &=& C_1^{\alpha\beta}\nabla_\alpha
      \hat{\cal R}_2^{\mu\nu}\nabla_\beta
      \hat{\cal R}_{3\,\mu\nu},\nonumber\\[1.5mm]
\overline{\Re_1\Re_2\Re_3}({21})
  &=& C_1^{\mu\nu}\nabla_\mu\nabla_\lambda
      \hat{\cal R}_2^{\lambda\alpha}
      \hat{\cal R}_{3\,\alpha\nu},\nonumber\\[1.5mm]
\overline{\Re_1\Re_2\Re_3}({26})
  &=& \nabla_\alpha\nabla_\beta C_1^{\mu\nu}
      \nabla_\mu\nabla_\nu C_2^{\alpha\beta}
      \hat{P}_3.
\end{eqnarray}
None of these structures contain a scalar curvature,
which makes them conformal invariant in a given
approximation.

The form factors $\overline{\Gamma}_i$ have several
integral representations. The most compact one
is so called $\alpha$-representation in which
the functions are given in terms of the integrals
\begin{equation}
\left<\frac{P(\alpha,\Box)}{-\Omega}\right>_3 =
\int_{\alpha \geq 0}
d\alpha_1\,d\alpha_2\,d\alpha_3\,
\delta(1-\alpha_1-\alpha_2-\alpha_3)\frac{P(\alpha,\Box)}{-\Omega},
\end{equation}
where
\[ \Omega =
\alpha_2\alpha_3\Box_1+\alpha_1\alpha_3\Box_2+\alpha_1\alpha_2\Box_3,
\]
and $P(\alpha,\Box)$ is an arbitrary polynomial in $\alpha$ with
$\Box$-dependent coefficients. In this representation the
form factors are
\begin{eqnarray}&&\overline{\Gamma}_{1}(-\Box_1,-\Box_2,-\Box_3) =
\left<\frac1{-\Omega}\Big(
   {1\over 3}
\Big)\right>_3
,\end{eqnarray}
\begin{eqnarray}&&\overline{\Gamma}_{2}(-\Box_1,-\Box_2,-\Box_3) =
\left<\frac1{-\Omega}\Big(
   {4\over 3}\alpha_1 \alpha_2 \alpha_3
\Big)\right>_3
+\frac13\frac{\ln(\Box_1/\Box_2)}{(\Box_1-\Box_2)}
,\end{eqnarray}
\begin{eqnarray}&&\overline{\Gamma}_{3}(-\Box_1,-\Box_2,-\Box_3) =
\left<\frac1{-\Omega} (
   2\alpha_1 \alpha_2)\right>_3
,\end{eqnarray}
\begin{eqnarray}&&\overline{\Gamma}_{5}(-\Box_1,-\Box_2,-\Box_3) =
\left<\frac1{-\Omega}\left(
   {1\over 9}\alpha_1
  +{1\over 9}\alpha_1 \alpha_2
  -{2\over 9}\alpha_1 \alpha_3
\right.\right.\nonumber\\&&\ \ \ \ \mbox{}\left.\left.
+{{\Box_1}\over {\Box_2}}\Big(
   {1\over 9}\alpha_2
  +{1\over 9}{{\alpha_2}^2}
\Big)
\right)\right>_3
+ {1\over {4 \Box_2}} - {{\Box_3}\over {24 \Box_1 \Box_2}}
,\end{eqnarray}
\begin{eqnarray}&&\overline{\Gamma}_{8}(-\Box_1,-\Box_2,-\Box_3) =
\left<\frac1{-\Omega}\left(
   {1\over 6}\alpha_2
  +{5\over 3}\alpha_1 \alpha_2
\right.\right.\nonumber\\&&\ \ \ \ \mbox{}
  -2{{\alpha_1}^2} \alpha_2
  -4\alpha_1 \alpha_2 \alpha_3
  +8\alpha_1 {{\alpha_2}^2} \alpha_3
\nonumber\\&&\ \ \ \ \mbox{}\left.\left.
+{{\Box_2}\over {\Box_1}}\Big(
  -{1\over 6}\alpha_1
  +{1\over 3}\alpha_1 \alpha_2
  +2{{\alpha_1}^2} \alpha_2
  +\alpha_1 \alpha_3
  +8{{\alpha_1}^2} \alpha_2 \alpha_3
\Big)
\right)\right>_3
,\end{eqnarray}
\begin{eqnarray}&&\overline{\Gamma}_{10}(-\Box_1,-\Box_2,-\Box_3) =
\left<\frac1{-\Omega}\Big(
   {1\over 3}\alpha_1 \alpha_2 \alpha_3
\Big)\right>_3
+ {1\over {270 \Box_3}}
\nonumber\\&&\qquad
- {{\Box_1}\over {540 \Box_2 \Box_3}}
,\end{eqnarray}
\begin{eqnarray}&&\overline{\Gamma}_{12}(-\Box_1,-\Box_2,-\Box_3) =
\frac1{\Box_1}\left[
\left<\frac1{-\Omega}\left(
   {8\over 3}{{\alpha_1}^2}
  -4{{\alpha_1}^3}
\right.\right.\right.\nonumber\\&&\ \ \ \ \mbox{}\left.\left.\left.
+{\Box_1 \over {\Box_2}}\Big(
   {1\over 3}\alpha_2
\Big)
+{\Box_1 \over {\Box_3}}\Big(
   {1\over 3}\alpha_3
\Big)
\right)\right>_3 \right.
\nonumber\\&&\ \ \ \ \mbox{}\left.
+\frac{\ln(\Box_1/\Box_2)}{(\Box_1-\Box_2)}
\Big({\Box_1 \over {3 \Box_3}}\Big)
+\frac{\ln(\Box_1/\Box_3)}{(\Box_1-\Box_3)}
\Big({\Box_1 \over {3 \Box_2}}\Big)\right]
,\end{eqnarray}
\begin{eqnarray}&&\overline{\Gamma}_{13}(-\Box_1,-\Box_2,-\Box_3) =
\frac1{\Box_1}\left[
\left<\frac1{-\Omega}\Big(
   2\alpha_1
\Big)\right>_3
+2\frac{\ln(\Box_2/\Box_3)}{(\Box_2-\Box_3)}\right]
,\end{eqnarray}
\begin{eqnarray}&&\overline{\Gamma}_{14}(-\Box_1,-\Box_2,-\Box_3) =
\frac1{\Box_3}
\left<\frac1{-\Omega}\Big(
   2\alpha_3
  -4{{\alpha_3}^2}
\Big)\right>_3
,\end{eqnarray}
\begin{eqnarray}&&\overline{\Gamma}_{16}(-\Box_1,-\Box_2,-\Box_3) =
\frac1{\Box_1}
\left<\frac1{-\Omega}\Big(
   {4\over 9}\alpha_1
  +{4\over 3}\alpha_2 {{\alpha_1}^2}
  -{4\over 9}\alpha_1 \alpha_3
\right.\nonumber\\&&\ \ \ \ \mbox{}\left.
  -{4\over 3}{{\alpha_1}^2} \alpha_3
\Big)\right>_3
+{1\over {6\Box_1\Box_2}}
,\end{eqnarray}
\begin{eqnarray}&&\overline{\Gamma}_{17}(-\Box_1,-\Box_2,-\Box_3) =
\frac1{\Box_1}\left[
\left<\frac1{-\Omega}\Big(
   2{{\alpha_1}^2}
\Big)\right>_3
+\frac{\ln(\Box_2/\Box_3)}{(\Box_2-\Box_3)}\right]
,\end{eqnarray}
\begin{eqnarray}&&\overline{\Gamma}_{18}(-\Box_1,-\Box_2,-\Box_3) =
\frac1{\Box_1}
\left<\frac1{-\Omega}\Big(
   2{{\alpha_1}^2}
  -8{{\alpha_1}^2} \alpha_2
\right.\nonumber\\&&\ \ \ \ \mbox{}\left.
  +8{{\alpha_1}^2} \alpha_2 \alpha_3
\Big)\right>_3
,\end{eqnarray}
\begin{eqnarray}&&\overline{\Gamma}_{19}(-\Box_1,-\Box_2,-\Box_3) =
\frac1{\Box_1}\left[
\left<\frac1{-\Omega}\Big(
  -4{{\alpha_1}^2} \alpha_2 \alpha_3
\Big)\right>_3
\right.\nonumber\\&&\ \ \ \ \mbox{}\left.
+\frac16\frac{\ln(\Box_2/\Box_3)}{(\Box_2-\Box_3)}\right]
,\end{eqnarray}
\begin{eqnarray}&&\overline{\Gamma}_{21}(-\Box_1,-\Box_2,-\Box_3) =
\frac1{\Box_1}\left[
\left<\frac1{-\Omega}\Big(
   8{{\alpha_1}^2} \alpha_3
  -16{{\alpha_1}^2} {{\alpha_3}^2}
\Big)\right>_3
\right.\nonumber\\&&\ \ \ \ \mbox{} \left.
-\frac23\frac{\ln(\Box_2/\Box_3)}{(\Box_2-\Box_3)}\right]
,\end{eqnarray}
\begin{eqnarray}&&\overline{\Gamma}_{24}(-\Box_1,-\Box_2,-\Box_3) =
\frac1{\Box_2}\left[
\left<\frac1{-\Omega}\left(
  -{5\over {54}}\alpha_2
  -{{23}\over {270}}\alpha_1 \alpha_2
  +{2\over 5}{{\alpha_1}^2} \alpha_2
\right.\right.\right.\nonumber\\&&\ \ \ \ \mbox{}
  -{4\over {15}}{{\alpha_1}^3} \alpha_2
  +{1\over {270}}{{\alpha_2}^2}
  +{{13}\over {270}}\alpha_2 \alpha_3
  -{1\over 5}\alpha_1 \alpha_2 \alpha_3
  +{4\over {15}}\alpha_1 \alpha_2 {{\alpha_3}^2}
\nonumber\\&&\ \ \ \ \mbox{}\left.\left.
+{{\Box_2} \over {\Box_1}}\Big(
  -{2\over {45}}\alpha_1
  +{1\over {45}}\alpha_1 \alpha_2
  +{1\over {45}}\alpha_1 \alpha_3
\Big)
\right)\right>_3
\nonumber\\&&\ \ \ \ \mbox{}\left.
+\frac{\ln(\Box_2/\Box_3)}{(\Box_2-\Box_3)}
  \left({{-\Box_2}\over {30 \Box_1}} \right)\right]
\nonumber\\&&\ \ \ \ \mbox{}
+{1\over {540\Box_2\Box_3}}
,\end{eqnarray}
\begin{eqnarray}&&\overline{\Gamma}_{25}(-\Box_1,-\Box_2,-\Box_3) =
\frac1{\Box_1}\left[
\left<\frac1{-\Omega}\left(
  -{{13}\over {135}}\alpha_1
  -{{56}\over {135}}\alpha_1 \alpha_2
\right.\right.\right.\nonumber\\&&\ \ \ \ \mbox{}
  +{{28}\over {45}}\alpha_1 {{\alpha_2}^2}
  +{{32}\over {45}}{{\alpha_1}^2} {{\alpha_2}^2}
  +{{16}\over {15}}\alpha_1 {{\alpha_2}^2} \alpha_3
\nonumber\\&&\ \ \ \ \mbox{}
+{{\Box_1} \over {\Box_3}}\Big(
  -{8\over {45}}\alpha_3
  -{{37}\over {135}}\alpha_1 \alpha_3
  +{{16}\over {45}}{{\alpha_1}^3} \alpha_3
  +{{11}\over {135}}\alpha_2 \alpha_3    \nonumber\\&&\ \ \ \ \mbox{}
  +{{28}\over {45}}\alpha_1 \alpha_2 \alpha_3
  -{4\over {45}}{{\alpha_2}^2} \alpha_3
  -{{16}\over {45}}\alpha_1 {{\alpha_2}^2} \alpha_3
  +{1\over {135}}{{\alpha_3}^2}
\nonumber\\&&\ \ \ \ \mbox{} \left.\left.
  +{{32}\over {45}}\alpha_1 {{\alpha_3}^2}
  -{{16}\over {45}}{{\alpha_1}^2} {{\alpha_3}^2}
  -{{32}\over {45}}\alpha_1 \alpha_2 {{\alpha_3}^2}
  +{{16}\over {45}}{{\alpha_2}^2} {{\alpha_3}^2}
\Big)
\right)\right>_3
\nonumber\\&&\ \ \ \ \mbox{}\left.
+\frac{\ln(\Box_1/\Box_2)}{(\Box_1-\Box_2)}
  \left( {{-2\Box_1}\over{15\Box_3}} \right)\right]
\nonumber\\&&\ \ \ \ \mbox{}
-{{1}\over {135\Box_1\Box_3}} + {{1}\over {270 \Box_2 \Box_3}}
,\end{eqnarray}
\begin{eqnarray}&&\overline{\Gamma}_{26}(-\Box_1,-\Box_2,-\Box_3) =
\frac1{\Box_1\Box_2}
\left<\frac1{-\Omega}\Big(
   4{{\alpha_1}^2} {{\alpha_2}^2}
\Big)\right>_3
,\end{eqnarray}
\begin{eqnarray}&&\overline{\Gamma}_{28}(-\Box_1,-\Box_2,-\Box_3) =
\frac1{\Box_1\Box_2}
\left<\frac1{-\Omega}\Big(
   {8\over 3}{{\alpha_1}^2} {{\alpha_2}^2} \alpha_3
\Big)\right>_3
\nonumber\\&&\ \ \ \ \mbox{}
+{1\over {135\Box_1\Box_2\Box_3}}
,\end{eqnarray}
\begin{eqnarray}&&\overline{\Gamma}_{29}(-\Box_1,-\Box_2,-\Box_3) =
\frac1{\Box_1\Box_2\Box_3}
\left<\frac1{-\Omega}\Big(
 {8\over 3}{{\alpha_1}^2} {{\alpha_2}^2} {{\alpha_3}^2}
\Big)\right>_3
.\end{eqnarray}

With the help of the algorithms described in \cite{IV}
the $\alpha$-representation can be converted to several
other integral representations: the spectral, Laplace and
explicit one.

The effective action (\ref{Tresult}),
(\ref{Aresult}), (\ref{Cresult})
is the desired new representation of the  original
effective action in the generic field model. In the
conformal invariant model this representation performes the splitting
of (\ref{result3}) in the anomalous and conformal
invariant parts, and the action (\ref{Tresult}) reproduces
the conformal anomaly via $W_{\rm A}$ {\it exactly}. Strictly
speaking, this conformal decomposition has a natural interpretation
only in conformal invariant theories with the properties (\ref{conftr}).
But, as it has been checked by direct calculations, the representation
(\ref{Aresult})--(\ref{Cresult}) is also valid for
a theory with a generic operator $H$.

\section{Conclusions}

In conclusion let us dwell on possible implications
of the obtained result. One of the prospects opening with
the knowledge of the conformal invariant part $\overline{W}$
(actually, the knowledge of the one-loop 3-vertex for a generic
field theory) might be utilized in extending and generalizing the 2D
results for correlation functions, operator products, etc. to 4D
and higher dimensions \cite{Osborn}.

The main field of application is, however, the theory of
quantum black holes and quantum gravitational collapse.
Here the major problem is a quantum backreaction which is
expected to be dominated by the Hawking radiation. The latter
arises in the presence of the BH horizon and is, apparently,
of the nonperturbative nature, since its flux and density are
inverse to the mass of a black hole, which is proportional
to the curvature magnitude
\[
\Re \sim {\cal M}, \qquad
<T_{\mu\nu}>_{\rm Hawking} \sim \frac{1}{{\cal M}^4}.
\]
Apparently, this would mean that a naive curvature expansion breaks down,
but its conformal version with the {\em exact} anomalous
part $W_{\rm A}$ could still be useful in view of the following
observation. The BH metric in the vicinity of the horizon
is conformally equivalent to the homogeneous spacetime
$R\times H^3$ with the spatial section $H^3$ of constant
negative curvature \cite{BFZ}
\[
ds^2|_{\rm horizon} \approx e^{2\sigma}(-dt^2+dl^2_{H^3}).
\]
Therefore the effect of this conformal factor can be taken into
account by the proposed method of the conformal decomposition
provided the constant spatial curvature can also be incorporated
in the calculation of the conformal invariant part of the
effective action $\overline{W}$.

\section*{Acknowledgments}

The authors are grateful to G. A. Vilkovisky for setting the
problem of converting the covariant curvature expansion into
the conformal basis and fruitful discussions.
We also thank A. Zelnikov for his assistance in calculations at
the initial stage of this project.
This work was supported in part by the Russian Foundation
for Fundamental Research Grant 93-02-15594, International
(Soros) Science Foundation and Government of the Russian
Federation Grant MQY300, and the European Community Grant
INTAS-93-493. One of the authors (A.O.B.) is also grateful for
the support of this work provided by the Russian Research
Project "Cosmomicrophysics".

\end{document}